# Correlated breakdown of carbon nanotubes in an ultra-high density aligned array

Shashank Shekhar, Mikhail Erementchouk, Michael N. Leuenberger, and Saiful I. Khondaker[*]

*NanoScience and Technology Center and Department of Physics, University of Central Florida, 12424 Research Parkway, Orlando, 32826, USA.*

**Abstract**

We demonstrate that in a densely packed aligned array of single walled carbon nanotubes, the breakdown of one nanotube leads to a highly correlated breakdown of neighboring nanotubes, thereby producing a nano-fissure. We show that the origin of the correlation is the electrostatic field of the broken nanotubes that produces locally inhomogeneous current and Joule heating distributions in the neighboring intact nanotubes triggering their breakdowns in the vicinity of the broken nanotubes. Our results suggest that the densely aligned array behaves like a correlated solid.

Due to their exceptional electronic properties including very high mobility, near ballistic conductance and resistance against electromigration, single-walled carbon nanotubes (SWNTs) are considered to be a promising building block for future digital and analog electronics [1-4]. Since it is difficult to control the chirality of individual nanotubes, arrays of SWNTs are becoming an important class of materials [5-27] as they can (i) average out inhomogeneities of individual tubes, (ii) provide larger on currents and (iii) reduce noise to provide higher cut-off frequency for radio frequency (RF) applications. Due to these advantages, there is a push to fabricate devices with massively parallel arrays of SWNTs for RF applications [5-7], field effect transistors [8-14], plastic electronics [8-9,15], display technologies [11, 15] and sensors [15]. However, since arrays contain a mixture of metallic and semiconducting nanotubes it is essential to selectively remove metallic nanotubes to obtain transistor properties [8-12, 16]. Such selective removal has been achieved for arrays containing up to 7 SWNT/µm assuming that the electrical breakdown of carbon nanotubes in air is due to Joule heating and oxidation [8]. It was also assumed that individual nanotubes can be broken without affecting the neighboring nanotubes and that only metallic nanotubes can be selectively removed leaving semiconducting nanotubes in the network. Detailed investigation was not done to see how many nanotubes were left behind. In addition, the authors did not check if their assumptions remained valid for arrays of density higher than 7 SWNT/µm.

A clear understanding of electrical breakdown of densely packed SWNT arrays is therefore essential for determining the fault-tolerance of nanotube arrays for electronic circuits. Since experimental and theoretical studies on individual SWNTs demonstrate that the breakdown is due to Joule heating and oxidation, which occurs at random places (at defect sites), a straightforward extrapolation of these studies to an array would suggest that the breakdown would occur at a random point inside each nanotube. Here we show in our experiments on arrays of nanotubes that while this straightforward extrapolation is valid only for large inter-nanotube separations (low density array), it does not hold for closely spaced nanotubes (high density array). We demonstrate that in a densely packed aligned array of SWNTs containing up to 30 SWNT/µm the breakdown of one of the nanotubes leads to a highly correlated breakdown of neighboring nanotubes, thereby producing a nano-fissure. Such a correlated breakdown does not occur if the SWNTs are not densely packed in the array implying that the correlation depends strongly on the inter-nanotube separation. We show theoretically that the correlated breakdown is due to the electrostatic field of the broken nanotubes that produces locally inhomogeneous current and Joule heating distributions in the neighboring intact nanotubes triggering their breakdowns in the vicinity of the broken nanotubes. Our results suggest that the densely packed aligned SWNT array behaves like a correlated solid and have strong implications in the future development of fault-tolerant nano-electronic circuits based on SWNT arrays.

To explore the correlated breakdown, the SWNT aligned array containing 1 SWNT/µm to 30 SWNT/µm were assembled from high quality aqueous solution via dielectrophoresis (DEP) between prefabricated Pd source and drain electrodes on Si/SiO$_2$ substrate. Figure 1a shows experimental setup for the SWNT assembly. Details of the alignment can be found in our previous report and supplementary material (see Fig. S1) [27, 28]. Figure 1b shows scanning electron micrographs (SEM) of a densely packed SWNT aligned array with a linear density of ~ 30 SWNT/µm. The current ($I$) - voltage ($V$) characteristics of this array (Figure 1c) show that $I$ increases rapidly to a few milli-Amperes and then decreases to zero as the $V$ was increased. Scanning electron microscopy (SEM) images taken afterwards show interesting nanoscale fissure in the aligned nanotube network (Figure 1d) occurred due to electrical breakdown of SWNTs.

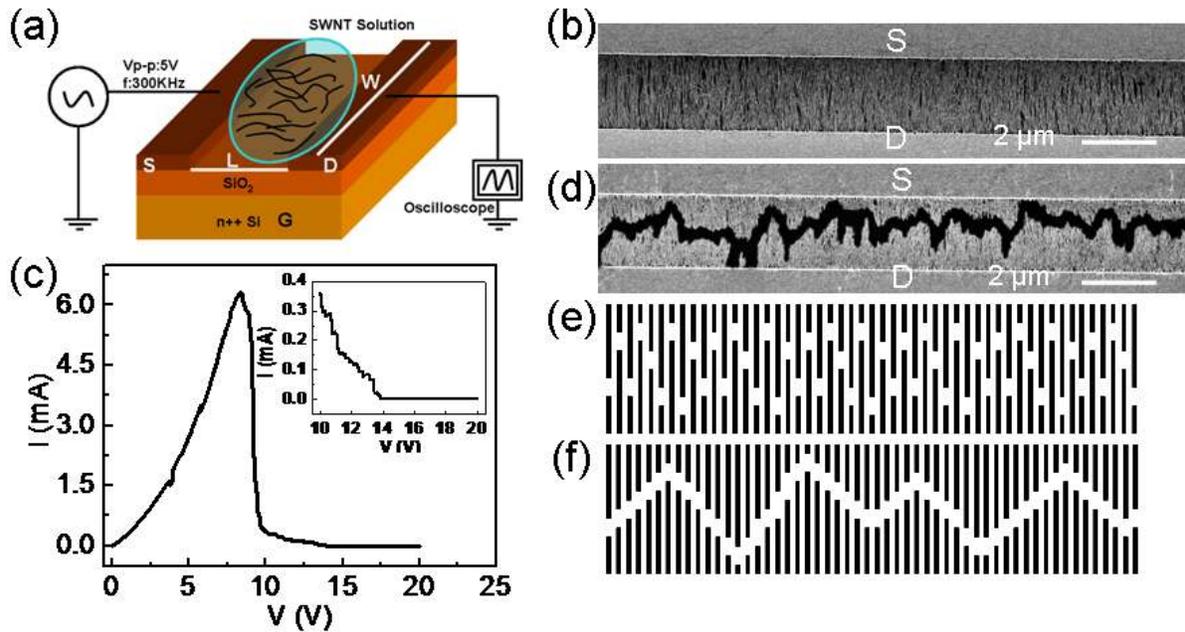

FIG. 1. Assembly and correlated breaking of a densely aligned SWNT array. (a) Schematic of the SWNT assembly setup. A small drop of high quality SWNT aqueous solution is placed between the Pd source (S) and drain (D) electrodes and an AC voltage is applied. (b) Scanning electron micrograph of an aligned SWNT network with ~30 SWNT/μm. L = 2 μm and W=25 μm. (c) shows the electrical breakdown process. Current is allowed to fall to ~ 0 (inset) by ramping DC voltage to more than 20 V. (d) Shows nano-fissure like continuous channel after breakdown. (e) shows the schematic of random breakdown of nanotubes. No pattern is formed. (f) Correlations in breakdown of adjacent nanotubes lead to the formation of a pattern.

The nano-fissure shows that the breakdown of individual nanotubes in the array is not independent of each other. If they were independent, we would see random breakdown of individual nanotubes giving rise to a random pattern similar to that shown in Figure 1e. This is because in an individual nanotube the breakdown occurs at the defect site due to Joule heating and oxidation at a critical temperature of ~600 $^0$C [17, 29-30] and one would expect that defects are randomly distributed. In contrast, our experiments demonstrate that the breaking of one nanotube affects the neighboring intact nanotubes in a correlated fashion forcing them to break in the vicinity of its broken region leading to the formation of a continuous channel similar to that shown in Figure 1f.

In order to examine what causes such a correlated breaking of SWNT, we performed breaking of arrays with varying SWNT density (Figure 2) with an average inter-nanotube separations ranging from 500 nm to 50 nm. At large inter-nanotube separation, the nanotube breaking occurs at random positions without any noticeable correlation between the breaking of neighboring nanotubes (Figure 2a-b). Figure 2c shows the breakdown of nanotubes when the density is ~ 8/μm, with an average inter-nanotube separation of 120 nm. Unlike Figure 2a-b, a pattern is becoming visible. The pattern becomes much clearer with increasing nanotube density (Figure 2d-e). It is clear that as the nanotubes become closer to each other, the breaking of one nanotube affects the neighboring tubes in a correlated fashion. Our results indicate that a noticeable correlation develops when the separation between nanotubes is less than 120 nm (array density >8 SWNT/ μm).

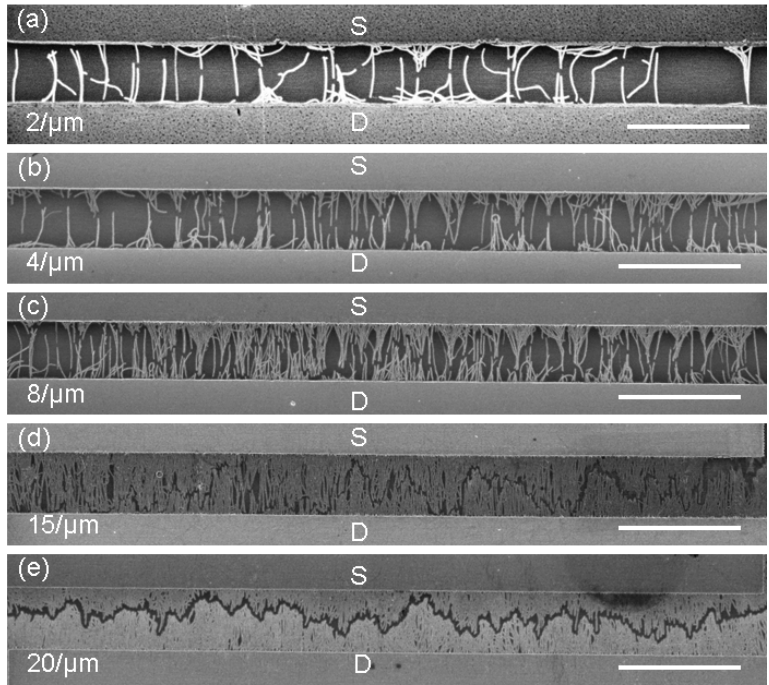

FIG. 2. Examination of correlated breakdown of SWNTs by varying the linear density of nanotubes. (a-b) shows the breaking of array with 2 and 4 SWNT/μm. The breaking of SWNTs are random and uncorrelated. (c) Array contains 8 SWNT/μm and a pattern seems to appear. (d-e) The correlated patterns become more pronounced at densities 15 and 20 SWNT/μm. The scale bars are 4 μm.

In order to further investigate the correlated nature of the breaking and how the nanotube breaking progresses, we performed electrical breakdown of SWNT arrays in three successive steps. Figure 3a shows the *I-V* characteristics of a representative array with 30 SWNT/μm. In the first two steps the current was allowed to drop approximately one order of magnitude from its peak value after which the voltage was swept back to zero. After the 3$^{rd}$ (final) breakdown, current drops to zero (Figure 3a inset). SEM image taken immediately after each breaking is shown in Fig 3b-d. After the first break, quite a few SWNTs are broken at random places all over the network (Figure 3b). However, no clear pattern is observed. After the 2$^{nd}$ breakdown of the same network, we estimate that more than 90% of the nanotubes are broken and a clear correlated pattern develops (Figure 3c). During the 3$^{rd}$ breakdown, the remaining nanotubes get broken and the nano-fissure gets wider (Figure 3d). Since the array contains both metallic and semiconducting nanotubes, our results demonstrate that the semiconducting nanotubes are not immune to correlated breakdown.

We now examine the cause for the correlated breakdown and formation of nano-fissure. The correlations could be regarded as an intrinsic feature of sufficiently dense networks or could be attributed to the transfer of heat produced during the burning of a nanotube to the neighboring tubes [5]. The latter mechanism, however, can be ruled out as it is not supported by the experimental data. Indeed, if this was the case then switching the bias voltage off after the first break (see Figure 3) would interrupt the formation of the correlated pattern and the nanotube network would proceed breaking at random positions after restoring the high bias voltage. The experiments, however, show that even after restoring the high bias voltage the breaking keeps its continuous correlated form (Figure 3d-e), which means that heat transfer cannot be responsible for the correlated breakdown. This suggests that the correlation is a manifestation of an intrinsic feature of a sufficiently dense network. We propose that the correlations are caused by the electrostatic field produced around the gap in the broken nanotube. While this field does not create a current, it alters the electron flow in the neighboring intact nanotubes thus affecting the distribution of the intensity of Joule heating. The most significant effect can be expected across the gap in the broken nanotube. This promotes the breaking of yet intact nanotubes near already existing breaks thus leading to the formation of a nano-fissure.

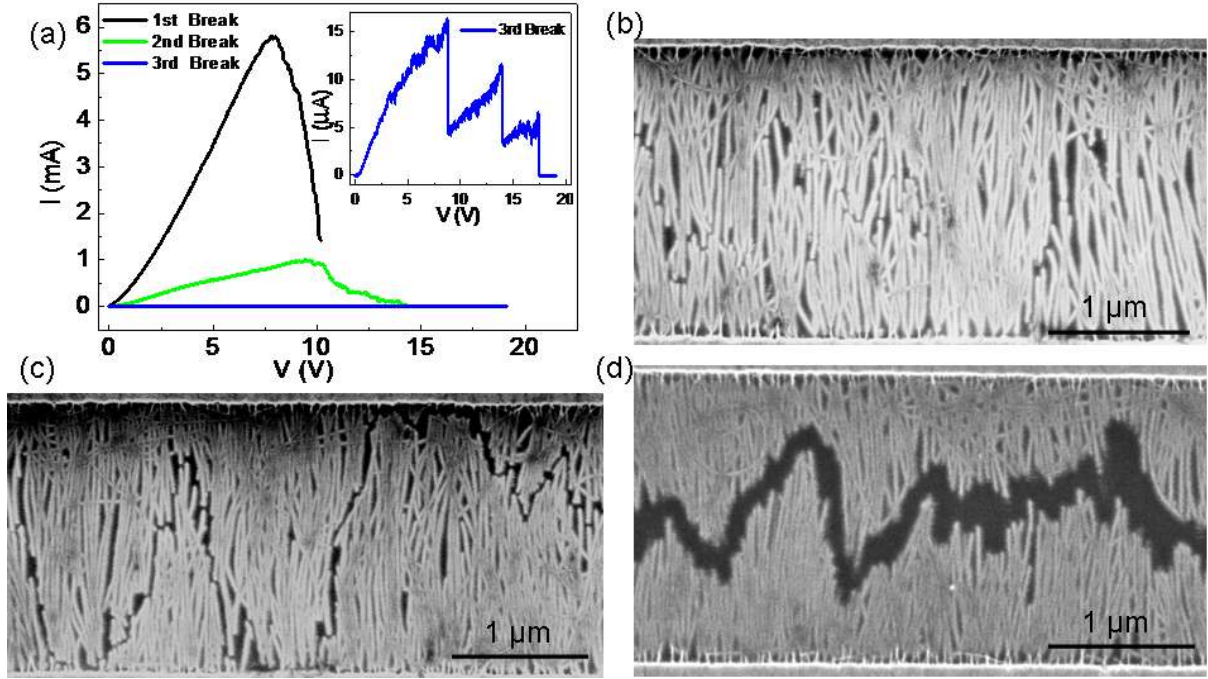

FIG. 3. The controlled breaking of an array with 30 SWNT/μm in three steps. (a) I-V curve demonstrating three successive electrical breakdown steps. V was swept back to zero after each step. Inset show I-V curve during the third step. (b) SEM image after the 1$^{st}$ breakdown. A small gap has appeared at random locations in the array of SWNT although no clear pattern can be seen. (c) SEM image after 2$^{nd}$ breakdown showing a continuous and clear pattern due to correlated breaking of SWNT. (d) SEM image after the final breakdown showing a wider continuous channel.

In order to show the mechanism in more detail, we consider a single nanotube that is broken at the point $z_0 = 0$ (see Figure 4a) and analyze its effect on the current distribution in a neighboring intact nanotube. Under the action of the external bias field the electric charge in the broken nanotube redistributes and creates an induced potential which breaks the cylindrical symmetry of the intact nanotube. Assuming that the scattering effect of the induced potential is small we find for the current density in the intact nanotube

$$j(z,\varphi) = |\chi_0(z,\varphi)|^2 j_0, \tag{1}$$

where $j_0$ is the density in the absence of the broken tube and $\chi_0(z,\varphi)$ is a solution of an equation of the Mathieu type (see the Supplementary material for the technical details) [28].

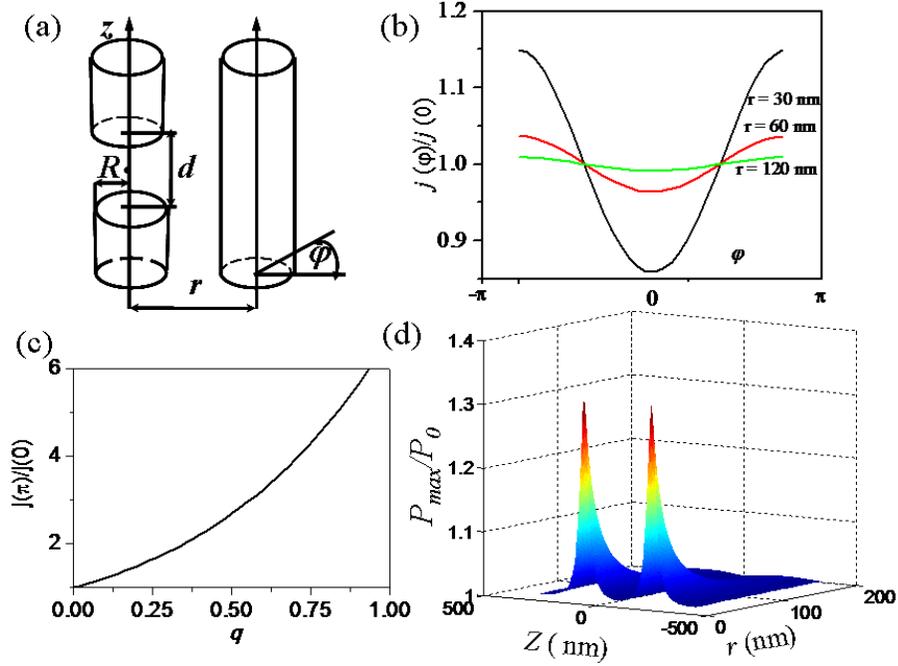

FIG. 4. The effect of the break on the intact nanotube. (a) Geometry of the problem. (b) The angular variation of the current density at $z = -d/2$ (the voltage $V=8$ V, the radius of the nanotube $R=1.5$ nm, the width of the gap $d=300$ nm) for different distances between the broken and intact nanotubes $r = 30$ nm, 60 nm, and 120 nm. (c) Dependence of the ratio of the maximum and minimum values of the current densities on $q$. (d) The ratio of the maximal power of Joule heating due to the current distribution $P_{max}$ and the power in the case when the broken nanotube is absent $P_0$ as a function of the distance between the broken and intact nanotubes and the position along the nanotube (other parameters are the same as for b).

Remarkably, the specific form of the spatial variation of the current density depends on the single parameter

$$q(z) = \frac{4mR^2}{\hbar^2} U_2(-z), \qquad (2)$$

which is a function of the system parameters, such as the geometrical sizes and the applied voltages. $R$ is the radius of the nanotube, $U_2(z) = eQRr(\rho_+^{-3} - \rho_-^{-3})$, $\rho_\pm = \sqrt{(z \pm d/2)^2 + r^2}$, $d$ is the width of the gap, $r$ is the inter-nanotube separation, and $Q$ is the induced charge, which is assumed to concentrate mostly near the edges of the gap. The induced charge is estimated by taking into account that $d \gg R$, which yields $Q = VR$, where $V$ is the applied external voltage.

Figure 4b shows the angular dependence of the current density for different $r$. Its variation can be characterized by the ratio of the densities along the opposite sides of the intact nanotube $j(\varphi = \pi)/j(\varphi = 0)$ plotted in Figure 4c as a function of $q$. In the shown range this ratio is well approximated by $j(\pi, z)/j(0, z) \approx \exp[2q(z)]$, which yields the explicit dependence of the current amplitude on the system parameters

$$j_{max} \approx j_0 \{1 + \tanh[q(z)]\}. \qquad (3)$$

The spatial variation of the current density leads to an inhomogeneous distribution of the intensity of the Joule heating $P(z,\varphi) \propto j^2(z,\varphi)$. In Figure 4d we show the maximal power $P_{max} \propto j_{max}^2$ as a function of the coordinate along the intact nanotube and its distance from the broken nanotube. It demonstrates that across the gap the Joule heating may noticeably exceed the one in the case of absent broken nanotube thus promoting breaking in this region. At the same time the effect drops fast with the inter-tube separation, implying that the correlations become significant only in sufficiently dense networks.

In conclusion, we demonstrated nano-fissure formation due to a correlated breakdown of sufficiently densed SWNT aligned array. The physical origin of the correlation is in the electrostatic field of the broken nanotubes that produces locally inhomogeneous current and Joule heating distributions in the neighboring intact nanotubes triggering their breakdowns in the vicinity of the broken tubes. Our results suggest that densely aligned SWNT behaves like a correlated solid and have strong implications in the future development of fault-tolerant nano-circuits using dense arrays of nanotubes. We expect such a phenomenon to be visible in other one-dimensional array systems invloving nanowires, graphene nanoribbons.

**Acknowledgements:** This work was partially supported by the US NSF under grant ECCS 0748091 (CAREER) to SIK, ECCS-0901784 to MNL and AFOSR under grant FA9550-09-1-0450 to MNL.

# Supplementary Material

# Correlated breakdown of carbon nanotubes in an ultra-high density aligned array

Shashank Shekhar, Mikhail Erementchouk, Michael N. Leuenberger, and Saiful I. Khondaker[*]
*NanoScience and Technology Center and Department of Physics, University of Central Florida, 12424 Research Parkway, Orlando, 32826, USA.*

\* To whom correspondence should be addressed. E-mail: saiful@mail.ucf.edu

**Alignment of Carbon Nanotubes of Controlled Density:**

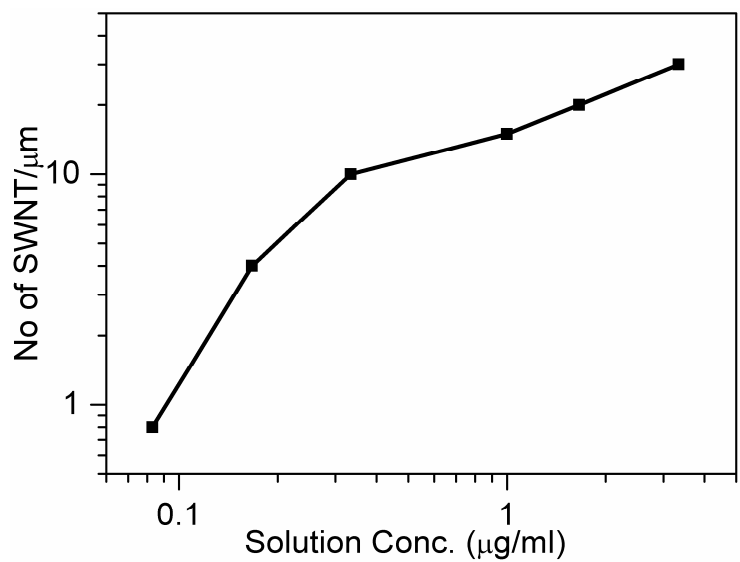

FIG. S1. Controlling the linear density of SWNT during DEP assembly. Number of nanotubes in the array was controlled by simply tuning the concentration of the SWNT solution for L=2 μm and W= 25 μm devices. V= 5 Vp-p, f=300 KHz and t=30 sec were used for the DEP assembly.

The SWNT aligned array of tunable density was assembled from a high quality aqueous solution obtained from Brewer science [1]. The solution has an original SWNT concentration of ~ 50 μg/ml and was diluted using de-ionized (DI) water to a desired concentration. The directed assembly of SWNTs at predefined electrodes was done in a probe station under ambient conditions via DEP. Prior to assembly, the electrodes were placed in oxygen plasma cleaner for 10 minutes to remove any unwanted organic residues on the surface. For the assembly, a small (3 µL) drop of the SWNT solution was cast onto the chip containing the electrode arrays. An AC voltage 5 Vp-p of frequency 300 KHz, was applied using a function generator between the source (S) and drain (D) electrodes for 30 seconds. The AC voltage gives rise to a time averaged dielectrophoretic force. For an elongated object it is given by $F_{DEP} \propto \varepsilon_m \, \text{Re}[K_f] \nabla E_{RMS}^2$, $K_f = \dfrac{\varepsilon_p^* - \varepsilon_m^*}{\varepsilon_m^*}$, $\varepsilon_{p,m}^* = \varepsilon_{p,m} - i\dfrac{\sigma_{p,m}}{\omega}$, where $\varepsilon_p$ and $\varepsilon_m$ are the permittivity of the nanotube and solvent respectively, $K_f$ is the Claussius-Mossotti factor, $\sigma$ is the conductivity, and $\omega = 2\pi f$ is the frequency of the applied AC voltage [2]. The induced dipole moment of the nanotube interacting with the strong electric field causes the nanotubes to move in a translational motion along the electric field gradient and align in the direction of the electric field lines. After the assembly, we determine the linear density of the SWNT from the SEM image. By simply varying the concentration, while keeping all other DEP parameters fixed we can reproducibly control the linear density of the nanotubes in the array from ~1 SWNT/µm to ~ 30 SWNT/µm. Fig. S1 summarizes the density of the nanotubes in aligned arrays vs. concentration of SWNT solution for channel length L= 2 and W= 25 µm device. Using a solution concentration of 0.083 µg/ml (600 times dilution) in L=2 µm device, we obtain ~1 SWNT/µm while a concentration of concentration of 3.3 µg/ml (15 times dilution) provided ~30 SWNT/µm.

**Theoretical formalism:**

In terms of the electron non-equilibrium Green's function $G^{-+}(\mathbf{r}_1, \mathbf{r}_2)$ the current density is written as [3]

$$j(\mathbf{r}) = \dfrac{e\hbar}{m}(\nabla_2 - \nabla_1) G^{-+}(\mathbf{r}_1, \mathbf{r}_2)\Big|_{\mathbf{r}_1 = \mathbf{r}_2 = \mathbf{r}}. \quad\quad [S.1]$$

In order to estimate the current density we employ the fact that we need to pay most attention to the relatively narrow region around $z = z_0$, where, in turn, we are mainly interested in the effect of breaking the cylindrical symmetry. Thus we write down the spectral representation

$$G^{-+}(\mathbf{r}_1, \mathbf{r}_2) = \sum_\mu G_\mu^{-+} \psi_\mu(\mathbf{r}_1) \psi_\mu^*(\mathbf{r}_2), \quad\quad [S.2]$$

where $\mu$ runs over the full set of quantum numbers characterizing the respective electron states, $\psi_\mu(\mathbf{r})$, in the nanotube, $G_\mu^{-+} = -2\pi i n_\mu \delta(\varepsilon_\mu - \varepsilon_F)$, where $n_\mu$ is the distribution function over the states with the energies $\varepsilon_\mu$ and $\varepsilon_F$ is the Fermi level. The electron wave functions $\psi_\mu(\mathbf{r})$ in this approximation satisfy the Schrodinger equation

$$-\dfrac{\hbar^2}{2m}\nabla^2 \psi_\mu(\mathbf{r}) + U_b(z,\varphi)\psi_\mu(\mathbf{r}) = \varepsilon_\mu \psi_\mu(\mathbf{r}), \quad\quad [S.3]$$

where $\varphi$ is the angle around the nanotube (see Fig. 4a, main text). The equation differs from the equation for a single tube by the additional external potential $U_b$, which is produced by the charge redistribution in the broken nanotube, leading to the electrostatic potential inside the neighboring intact nanotube as shown in Fig. S5a. Assuming that the excessive charge is mostly localized near the gap edges and keeping only first two terms in the respective Fourier expansion we present $U_b$ in the form

$$U_b(z,\varphi) \approx U_1(z) - U_2(z)\cos(\varphi), \qquad [S.4]$$

where $U_1(z) = eQ(\rho_+^{-1} - \rho_-^{-1})$, $U_2(z) = eQRr(\rho_+^{-3} - \rho_-^{-3})$, $\rho_\pm = \sqrt{(z \pm d/2)^2 + r^2}$, $r$ is the distance between the nanotubes, $Q$ is the charge induced near the edges of the gap in the broken nanotube and $R$ is the radius of the tube. We estimate $Q$ by taking into account that the width of the gap is much larger than the radius of the tube. We find $Q = VR$, where $V$ is the applied external voltage.

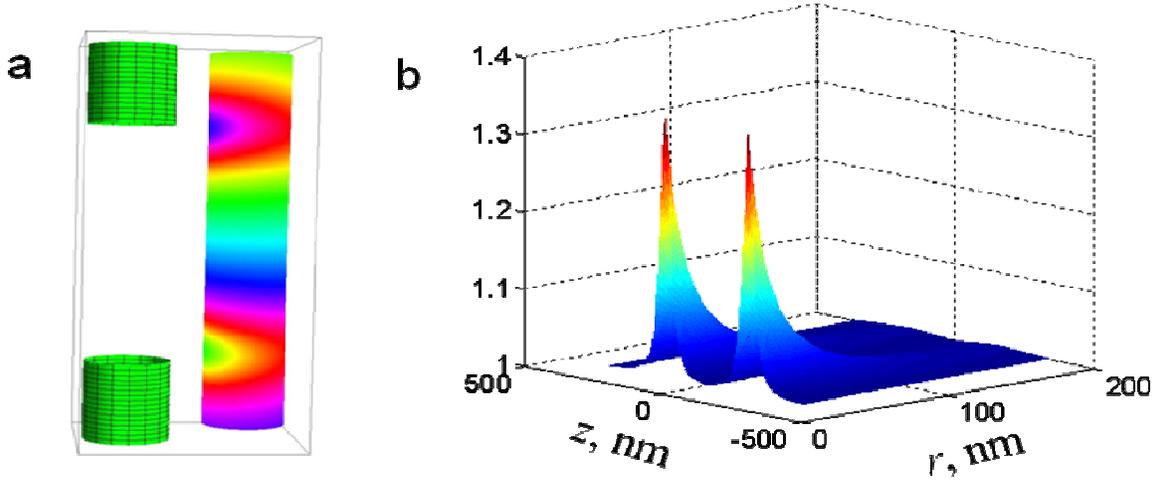

FIG. S2. (a) Electrostatic potential produced by the broken nanotube inside its neighboring intact nanotube. (b)The ratio of the current densities at the opposite sides of the intact nanotube as a function of the distance between the broken and intact nanotubes and the position along the nanotube.

We assume that $U_b$ does not lead to the essential reconstruction of electron eigenstates. Thus, we can present the solution of Eq. [S.3] in the form

$$\psi_\mu = \left(\psi_\mu^{(0)} + \psi_\mu^{(1)}\right)\chi_0(z,\varphi), \qquad [S.5]$$

where $\psi_\mu^{(0)}$ is the electron wave-function in the absence of the external perturbing field and $\psi_\mu^{(1)}$ is a small correction accounting for scattering due to the induced potential. The angular modulation due to the perturbation is described by $\chi_0(z,\varphi)$, which is the periodic solution of the equation

$$\frac{\hbar^2}{2mR^2}\frac{\partial^2 \chi_0}{\partial \varphi^2} + \left[\varepsilon_M(z) + U_2(z)\cos(\varphi)\right]\chi_0 = 0. \qquad [S.6]$$

This equation is of the Mathieu type [4] with $z$ being the parameter. The characteristic values admitting the periodic solutions form the discrete sequence of real numbers, out of which we are interested in the lowest one, $\varepsilon_M(z)$. The specific form of the solutions is governed by the parameter of the Mathieu equation $q$, which in terms of the parameters of the problem is expressed as

$$q(z) = \frac{4mR^2}{\hbar^2} U_2(-z). \tag{S.7}$$

Thus the dependence of the angular variation of the current density on the coordinate along the nanotube, applied voltage, the distance between the nanotubes and so on is combined into the single parameter $q$. With this regard the strong dependence of $q$ on the radius of the nanotubes, $q \propto R^4$, should be mentioned.

Using representation [S.5] we find from Eqs. [S.1] and [S.2]

$$j(z,\varphi) = |\chi_0(z,\varphi)|^2 j_0, \tag{S.8}$$

where $j_0$ is the current density in the absence of the broken tube in the neighborhood. In order to illustrate the dependence of the angular variation of the current density on the parameters of the problem we plot in Fig. 4 c in the main text the ratio of the current densities on the opposite sides of the nanotube $j(\varphi = \pi)/j(\varphi = 0)$ as a function of $q$. In the shown range this ratio can be approximated by

$$j(\pi)/j(0) \approx \exp[2q], \tag{S.9}$$

which yields for the amplitude of the current $j_{max} \approx j_0 \tanh[q]$. As can be seen from [S.7] the maximum variation is reached across the gap edges, which is illustrated by Fig. S2 b.

**References**

1. Brewer, http://www.brewerscience.com/.
2. M. Dimaki, P. Boggild, Nanotechnology **15**, 1095 (2004).
3. M. A. Zagoskin, *Quantum theory of many-body systems: techniques and applications* (Springer, New York, 1998).
4. M. Abramovitz, I. Stegun, *Handbook of Mathematical Functions* (Dover, New York, 1970).